# How Large Language Models (LLMs) Extrapolate: From Guided Missiles to Guided Prompts

**Author**

Xuenan Cao,
Assistant Professor of Cultural Studies, Chinese University of Hong Kong
xuenancao@cuhk.edu.hk

**Abstract**

This paper argues that we should perceive LLMs as machines of extrapolation. Extrapolation is a statistical function for predicting the next value in a series. Extrapolation contributes to both GPT's successes and controversies surrounding its "hallucination" (Alkaissi and McFarlane 2023). The term "hallucination" implies a malfunction, yet this paper contends that it in fact indicates the chatbot's efficiency in extrapolation, albeit an excess of it. This article bears a historical dimension: it traces extrapolation to the nascent years of cybernetics. In 1941, when Norbert Wiener transitioned from missile science to communication engineering, the pivotal concept he adopted was none other than "extrapolation." Soviet mathematician Andrey Kolmogorov, renowned for his compression logic that inspired OpenAI, had developed in 1939 another extrapolation project that Wiener later found rather like his own. This paper uncovers the connections between hot war science, Cold War cybernetics, and the contemporary debates on LLM performances.

## 1. Introduction

You have just entered a query, and your chatbot has begun generating output—a sequence of words and punctuations, appearing one by one, traveling down a path from left to right. Each token, be it a word, a punctuation, or a phrase, is a message. The "message of any medium or technology is the change of scale or pace or pattern that it introduces into human affairs" (McLuhan 1964, 1). Here, the message of the chatbot conveys the correlation between words and phrases, i.e., insights that a pretrained model had learned from an unknown corpus of texts whose sheer size has drawn serious concerns in the critical AI community (Bender *et al.* 2021). Based on how well the pretrained model has been fine-tuned to perform specific tasks, the model can predicts the next most likely token in a series of units. If you had just thought of one word, what would be the most probable next word that would come to you? Your chatbot knows better. It generates a sequence of symbols one element at a time, using each previous output word as additional input for computing the next symbol, aligning not only words and sentences, but also paragraphs (Vaswani *et al.* 2017). The series initiates with one word, followed by another, and eventually draws to a closure when the model's "stop" condition is met. Between the beginning and the end, we find elements, each of which is associated with a value in the deeply layered neural net. Series of elements unfold in time like darts flying through successive points in the air. There could had been many paths for each dart, but once the first is thrown and the sequence is set in motion, the trajectories of the followings can be predicted and calculated as time series. Every value within the time series maintains a statistical correlation with other units.

When you ask for a reference on a specific topic and ChatGPT gives you made-up citation, the LLM behind it fabricates this serial output. The output is nothing more unreal than what was statistically probable in a series: the likely make-up of an article, including its title, an actual name most likely to be associated with it, the journal that would perhaps publish it if it were real, and a pseudo DOI. The chatbot has extrapolated.

In just a few months after the launch of GPT for public use, medical scientists proposed with good reasons the term "Artificial Hallucination," a phrase intended to pathologize this commodity bot in its infancy (Alkaissi and McFarlane 2023; Beutel *et al.* 2023). The terminology refers to a widely discussed phenomenon of chatbot making up fake references and posing a real danger to scientific research. The term "hallucination," laden with a pathological connotation, suggests that something within the system is amiss. In an academic landscape increasingly oriented toward digital humanities and the steady importation of quantitative tools into virtually all fields, this impression that "it does not work" tends to serve as a moment of retribution: by spotting the chatbot's shortcomings, we, scholars and educators, can find reassurance against AI's encroachment into our domains. So, we are on board when boycotting GPT-written codes, projects, and summaries. We remain vigilant in the face of chatbots' overconfidence, surmising that AI developers may have misleadingly hyped their bots' excellence. The phrase "hallucinated text" entails implausibility and nonsense, but, from the vantage point of Large Language Models (LLMs), false references are in fact quite plausible sequences of words that are not exact replications of facts in its training set, but still highly sensible outputs (Klein 2023). How might we make sense of functional behavior?

This paper argues that artificial hallucination has another name: extrapolation. Extrapolation indicates that a machine learning model must leap beyond the boundaries of its training data, i.e., generate something novel that was not in the training data. According to the classic textbook in artificial intelligence, this capability "to adapt to new circumstances and to detect and extrapolate patterns" defines machine learning (Russell and Norvig 2010, 3). What makes the chatbot fabricate a non-existent citation is also what enables GPT to profit and function. An extrapolating model means it is working, and *working too well* according to its design. For instance, extrapolation is tied to your chatbot's ability to entertain you with a poem. It had to learn the poetic forms and then extrapolate from its data to create new output. Extrapolation also helps a chatbot avoid plagiarism. By extrapolating beyond the training data, a model can generate an output without exposing it source. As a result, we cannot tell if the model has been trained on, say, the archive of *The New York Times* or any other copy-righted materials available online. This evasion is a desired type of "intentional opacity" (Burrell 2016) protecting the AI developers from potential copyright claims to their profits from the product.

Extrapolation as a functional feature of LLM is the first argument of this paper. The other argument of the paper is a historical one, showing that extrapolation as a concept has roots in the development of cybernetics. Unbeknownst to many in both humanities and computer science, extrapolation can be traced back to a specific historical juncture when cybernetics was still just on the horizon in Norbert Wiener's intellectual enterprise. In 1941, Norbert Wiener explored how he might use the principles he had developed for predicting missile trajectories in other fields. The projection of a missile, a problem that occupied the mind of many wartime scientists, was perceived as a "time series," a collection of points that appear sequentially in time. In time series analysis, the position of a missile depends on its previous location. The placement of a word, too, depends on its preceding words. In Wiener's mind, a sequence of words aligned with the notion of time series. "The message to be transmitted is represented as some sort of array of measurable quantities distributed in time" (Wiener 1964, 2). Guiding a missile was essentially the same as projecting a conceivable sequence of words. Wiener's representational endeavor of encapsulating the progression of words as temporal sequences held vast potential. Throughout 1941, Wiener transposed the mathematical formulation employed to predict missile trajectories onto the probable course of units within sequences of messages handled by apparatus such as the telephone.

The mathematical term Wiener used to describe the predictive function for time series analysis is extrapolation, found in his work titled *Extrapolation, Interpolation, and Smoothing of Stationary Time Series with Engineering Applications* (henceforth *Extrapolation*). The initial submission of this work to the Office of Defense in 1942 and its subsequent reprintings for public use in the 1950s and 1960s had an evolving context to them. As I will detail in the history section, this highly technical work was crucial in the phase immediately preceding Wiener's turn from artillery science to his more ambitious cybernetic project.

Coincidentally, extrapolation was also a key concept in the works of Soviet mathematician Andrey Nikolayevich Kolmogorov, whose studies bore an almost identical title, "Interpolation and Extrapolation of Stationary Series" (1939). [1] To this parallel line of thinking by his younger Soviet counterpart, Wiener showed grudging deference. Through this convergence, he recognized Kolmogorov as an intellectual peer when Kolmogorov faced pushback from other advocates of Soviet cybernetics in the 1940s USSR (Peters 2016, 40-41).

Nowadays, Kolmogorov is a term found in the textbooks on learning theory in computer science. He had inspired advanced methods in deep learning that eventually led to ChatGPT. His name has appeared in countless papers because it denominates algorithms used for compression. The size of the training data necessitates compression, which then results "in a loss of fidelity in the final statistical model" (Hiller 2023). Compression algorithms distill complex series into simpler forms by finding the probability distribution of input, thus allowing fewer digits to encapsulate a series containing more symbols. Lossy compression is necessary for reducing any social realities into calculable sets. A dataset first turns real life data into accessible points, each capable of being "addressed" (Dhaliwal 2022) independently or as a whole, then compression selects its patterns to obliterate other details, like a Piet Mondrian abstraction of the Manhattan skyline into blue, red, and yellow stripes. What Kolmogorov achieved through computation was a promise to mathematically refine of complexity by abstraction and "reductionism" (Vigini 2023). The reduced world can be stored as dots and dashes on perforated tapes and punched cards, later as digits, discontinuous yet whole. To jump the discontinuities ingrained in data to recreate that wholeness is the goal of generative AI (Chiang 2023). The ability to perform the wholeness by discontinuous data comes from a model's capacity to interpolate and extrapolate.[2]

---

[1] There had been several translations of this same work, one of them is "Interpolation and Extrapolation of Stationery Random Series" (Kolmogoroff 1941). The citation that Wiener himself used was a slightly earlier source (Kolmogorov 1939), without including the word "random" in the title.

This article argues that we cannot understand how the chatbots function without understanding the much longer history of Wiener's and Kolmogorov's extrapolation, even though this mathematical concept has only very recently become a topic of interests in computer science. Responding to the call to reintroduce interpretation into the study of AI (Wachter *et al.* 2017; Petit 2023), this paper shows how we can use extrapolation as a pivotal perspective into the ongoing discussion of LLMs. I will first retrace Wiener's steps in developing this concept, unearthing the connections between extrapolation in wartime science and its subsequent Cold War era prominence in mathematicise (not, surprisingly, in computer science). Then, I will discuss the recent use of extrapolation as a contentious concept in computer science community. The two parts of the paper, one historical and the other conceptual, were simmered in distinct cultural and political contexts. Yet they convey a shared message: comprehending the workings of LLMs demands an appreciation of their extrapolation capabilities.

---

[2] Interpolation and extrapolation are the same statistical method. The difference is that interpolation means the estimation of value within the range of known values, whereas extrapolation means estimation beyond the range.

## 2. Extrapolation, A History

### From Missiles to Word Projectiles

To understand what extrapolation meant in a more distant era, before "cybernetics" and "information" had become words of interest, it is important to sketch the concerns in an older discipline - that of statistics. In the mid to late twentieth century, a snowballing number of mathematicians turned to statistics for wide-ranging military applications. A classic problem is time series analysis, i.e., predicting the next event appearing in time. A time series is a collection of continuous or discrete data over time. In the war against fascism, time series analysis had gone in many directions, from calculating the chance of a certain area being hit by a bomb to predicting the trajectory of a missile. Probability solved the problem of controlling seemingly random warring worlds.

The appeal of using this man-made concept of probability for developing military strategies and tactics on a battlefield was obvious. What was not so obvious was its application in communication. One notable figure who had taken a singular move on this path was Norbert Wiener. In *Extrapolation,* first submitted as a classified document to the US Defense Office in 1942, Wiener presented a crucial step towards what later become known as cybernetics: integrating communication engineering and time series computing (Wiener 1942/1964).[3]

To Wiener, a time series can be a projectile trajectory. His missile guidance algorithms predicts the trajectory of a ballistic unit. But what else could a time series be? In Wiener's particularly imaginative moves, he envisions the projection of messages unfolding in time like a missile trajectory. "While one does not ordinarily think of communication engineering in the same terms, this statistical point of view is equally valid there" (1942/1964, 4). Throwing a word, for Wiener, is also a prediction task. Transformation carried out by electrical means is "in no way essentially different from the operations…carried out…by statistician…and computing machines" (1942/1964, 2). According to Wiener, a message is a time series "subjected to transmission by an apparatus which carries it through a succession of stages (1942/1964, 2). A time series can be translated into signals, i.e., a different time series, with the help of a transmission apparatus, a black box.[4] Each black box transformation of a time series becomes a new time series.

---

[3] The texts I quote from this publication have been published in 1942. If contents have been quoted from its more recent reprint, such as the one published by MIT Press in 1964, I will note it otherwise.

Wiener sees the statistician in him fit for dealing a large quantity of time series and their transformations. "No apparatus for conveying information is useful unless it is designed to operate, not on a particular message, but on a set of messages, and its effectiveness is to be judged by how it performs on the average on messages of this set." (1942/1964, 4) The "average effect" is how Wiener judges a set of time series, not "effect in any particular case" (1942/1964, 4). To compute the distribution of not just one but a set of time series, what he calls "ensembles of time series" (1942/1964, 4), Wiener recreats text transmission in the imagery of a textual artillery bombardment in what an evolving "cognitive assemblage"[5] of humans and machines. By analyzing clusters of words and how the landing of one word would change the overall landscape, statistical techniques can make informed predictions about tokens to follow. A transmission apparatus tosses out a word like a jet ejecting a bomb. The jet helps the pilot predict where the bomb will land. Likewise, the model can learned how that word has been tossed into a cluster of words and correlates with each other. By extending this representational endeavor, Wiener invents a new science that combines control and message transmission, later known as cybernetics.

---

[4] Petrick (2019) writes a long history of the development of Blackbox as a wartime transmission technology. Galison (1994) also mentioned the use of black-speckled boxes that encase transmission desire.
[5] Here, I am with Katherine Hayles (2017; 2019) on the point that machine thinking does not need to be autonomous or even conscious for it to count as cognition. Yu-Shan Tseng (2023) has recently argued that despite the blackbox nature of many machine learning models, we can still comprehend urban AI landscape through seeing the distribution of AI in urban assemblage. The idea of assemblage is crucial here for understanding the related nature of distributed intelligence among humans and machines.

In this prediction task, the key mathematical formulation is extrapolation. "Let us now consider some of the things which we can do with time series or messages. The simplest operation which we can perform is that of extrapolating them" (1942/1964, 9). By this point, the report consists of the various methods aimed to aid the extrapolation of time series: "A…method of *extrapolating a time series into the future* is judged by the probability with which it will *yield an answer correct within certain bounds*" (1942/1964, 4). Here, accuracy is not the point; approximation is. Using any training set, we can measure the extent of extrapolation and know if the model can "yield an answer correct within certain bounds" (1942/1964, 4). Wiener wants to ensure that extrapolation remains a useful function for measuring accuracy, even if errors are inevitable. In the context of a chatbot making up references, the accuracy of the output is likewise not absolute. We might ask GPT to suggest a book on a certain topic and end up getting a fiction: it is plausible that a certain author might have written a book with a particular title that fits the query description. This book will likely exist in the real world, even though it might not exist now.

When the model extends its prediction beyond its training data, it extrapolates; otherwise, it interpolates. Both cases are considered mathematically accurate within a certain bound of errors. The difference is that, as Wiener points out, extrapolation "may be applied whenever an ideally desirable linear operation on a statistically uniform time series is not strictly realizable" (1942/1964, 23). If there was a realizable output already found in the past series, it means the model only needs to interpolate. If a scenario is not realizable, the model must extrapolate into new territories. In other words, the output might not be determinable completely in terms of its past (1942/1964, 54-55). When extrapolation is involved, as Wiener articulates, the model possesses the capacity to "introduce new information" and locate "singular" cases (1942/1964, 55).

Extrapolation, a function that suggests a model must generate outputs beyond its training set, could be considered open enclosures capable of transforming and emitting lines of flight - lines so unlikely, and yet seemingly inescapable as an output for a particular query, as if a word that knows its course before it was thrown. The flight path was at first quite literal in the case of missile guidance. Deleuze's notion of "line of flight" becomes particularly illuminating here. The term denotes an escape route from constraining structures. The trajectory of a message, a creative escape from the convex enclosure of the training data, takes on additional meaning. Rather than adhering to a fixed, predetermined path, an extrapolated message extends into the unknown, although it might not suggest an absolute escape from a territory that leads to new emergence and future singularities.[6]

### From Hot War Science to Cold War Publicity

The epistemological jump from missile to word projectile is not a purely intellectual shift. Although Wiener's work on time series enhanced missile guidance systems during the hot war, it was the subsequent period of the Cold War that propelled the popularization of Wiener's extrapolation project in different directions. During the war, Wiener's scientific work had seemed a heroic contribution to the defense against the fascists. After the dropping of the atomic bomb, a different fashion set in. Even though the tragedy of Pearl Harbor still weighed on the minds of many Americans, the atomic bomb weighed more on the conscience of scientists.

Wiener was troubled by his conscience, too, evidenced by his letter "A Scientist Rebels" (1947, 46). It was published in January 1947 in *The Atlantic.* Prior to this letter, the tone was already set by a series of images and journalistic accounts showing the catastrophic impact of the bombs on the civilians of Japan. In an article titled "Hiroshima," published in *The New Yorker* in August 1946, the reporter John Hersey (1946) traced the lives of six residents in this historic city, moments before the explosion of the bomb. The essay also reports how Japanese scientists strove to collect information about the bomb's explosion height and its uranium content to determine the range of radiation—a crucial step for cleaning up the mess. But such details were kept confidential by the US. In addition, it was known that the US had made no attempt to declassify or to ban further use of the weapon. The mood of this period was that was lament. It was time to rethink the influence of the US victory.

---

[6] Deleuze and Guatarri conceptualized rhizomes made of lines, and the most extreme escape routes, i.e., a line of flight, is pure abolition, destruction and deterritorialization. See Introduction on Rhizome in Deleuze and Felix (1987). Many books and articles (e.g., Savat 2013, 63-82; Sampson 2017; Murphie 1996) have been published on similar topics, discussing how Deleuze and Guattari's concepts of assemblage can be used in thinking about cognition of a different kind. Since this thread is outside the scope of the historical argument here, I have left this discussion out of the main text.

Wiener's letter "A Scientist Rebels" responds to a request Wiener had received from a research scientist of an unnamed aircraft corporation. "Sir: I have received from you a note in which you state that you are engaged in a project concerning controlled missiles and in which you request a copy of a paper which I wrote for the National Defense Research Committee during the war." (46) "A paper" refers to works conducted during the war but had not been reprinted for public use. Wiener saw no reason to make such work available to the aircraft corporation because "the practical use of guided missiles can only be used to kill foreign civilians indiscriminately" and he no longer "desire to participate in the bombing or poisoning of defenseless peoples" since US had already won two years ago (46).

Like him, other scientists grappled with the increasingly burdensome moral implications of their research. *The Atlantic* has published other articles featuring the indignation of esteemed scientists concerning the US's strategic stance. Among these voices was Albert Einstein, who penned an essay "Atomic War or Peace" published in November 1947. Einstein decried the US government's inaction in denouncing the use of atomic power, especially when this country, by 1947, had "the monopoly" over it (30). "[Americans] may believe it is superfluous to announce publicly that they will not a second time be the first to use the atomic bomb…This country has been solemnly invited to renounce the use of the bomb—that is, to outlaw it." (30) The US, much to Einstein's indignation, "has declined" to do so unless the world accepts US's proposed "terms for supernational control" (30). This quest for acknowledgment of supremacy struck many scientists like Einstein as unnecessarily provocative. The insistence on withholding a ban on the bomb until the USSR accepted the US' claim to supremacy was, in Einstein's view, nothing short of "the American Failure" (31).

This was the social atmosphere in which Wiener must demonstrate his awareness of his political surroundings and consider steering clear of any indignity. The patriotic motif could no longer justify his military contribution to war technologies. In his letter "A Scientist Rebels," Wiener writes that scientists must account for the disasters that they created out of their lab work: "The policy of the government itself during and after the war, say in the bombing of Hiroshima and Nagasaki, has made it clear that to provide scientific information is not necessarily an innocent act." (46) He called other scientists to make their independent decisions, while he himself attempted to turn his entire scientific enterprise into something as remote from war as possible.

Representing a string of words as a time series provided an ideal outlet for Wiener's talent, appealing to the scientific community who sought to reshape their images after the war and in line with the US public relations efforts. *Extrapolation* was first prepared as a classified document to the US National Defense Research Committee in 1942 and printed after the war, in 1949, for general use. The small volume was then reprinted several more times, each with a higher status in cybernetics. Not all of Wiener's war-time science works were published for later use, or at least not quickly and with so many reprints. It needs to be added that *Extrapolation* is a highly technical work hardly meant to be consumed by a non-specialist reader. Despite its high level of sophistication, the reprinting of it in 1949 for public use served a different function: it pivoted attention away from the hideous crime of the atomic bomb.

The publication of *Extrapolation* presented an opportunity for the state to save its reputation during the Cold War, diverting public attention to civilian use of technologies. In the preface to the 1964 reprint of *Extrapolation* by MIT Press, he said, "the present monograph represents one phase of the new theory pertaining to the methods and techniques in the design of communication systems" (1942/1964, v). There was no reference to how missile science was a part of Wiener's conceptual development. To calibrate the predictive capacity was the art of the highest kind. If, as Galison (1994) argues in his article "The Ontology of the Enemy: Norbert Wiener and the Cybernetic Vision," "the system of weaponry and people that Wiener had in mind was predicated on a picture of a particular kind of enemy…[that] was neither invisible nor irrational," this article shows that there was a turn away from this enemy ontology in Wiener's later work, or at least some efforts to steer clear from his "enemy" thinking, given the moral pressure from his contexts (229). Although Wiener focused his attention on the prediction of Nazi planes that throw projectile weapons, in the years immediately after the war, Wiener had tried to shift the public attention on him towards the civilian use of his earlier research. It is in this crucial turn in his work that *Extrapolation* became an important text.

**A Minor Scene**

Wiener wrote that *Extrapolation*'s purpose was to demonstrate "a complete natural methodological unity" (1942/1964, 1). Wiener noted that despite the similarities between time series analysis and communication problems, the methods used in those two fields had little in common, not until *Extrapolation*. But there was a minor scene that complicates

Wiener's claim to novelty. As Wiener submitted his report for printing in 1942, he was appalled that similar works had already been done by some Soviet mathematician called Andrei Nikolaevich Kolmogorov. A "Mr. I. E. Segal of the Princeton Graduate School" (1942/1964, 59) brought Kolmogorov's work to his attention, but not at any point important for Wiener's intellectual progress in this process, as Wiener hinted. Only at "the Christmas meeting of the American Mathematical Society in 1941" (59), after completing most of his extrapolation research, he became aware of this other work. Although, very cautiously, Wiener also acknowledged that a "Professor Fuller at Brown University" who had mentioned Kolmogorov's work "at an earlier period" from a "casual conversation" (59). This minor scene occupies only one footnote in Wiener's *Extrapolation.* Wiener cites Kolmogorov's paper of almost the same title "Interpolation and Extrapolation of Stationery Sequences," first published in French in 1939 in Bulletin de l'Académie des sciences de l'URSS, then printed again in Russian 1941 in Izvestiâ Akademii nauk SSSR, and then in German.[7] Wiener's citation of Kolmogorov in the footnote also appears to be incorrect, as he puts the publication date of the French work in the year 1941, not 1939. In his preface to the 1964 reprint, Wiener had corrected the mistake.

This was perhaps the first instance that begin the many future connections between Soviet mathematics and US cybernetics during the Cold War. Initially however, Wiener emphasized that Kolmogorov's results did not affect the progress of his independent thinking, even though Kolmogorov's publication came at least two years prior to his own:

> The author wishes to comment on the historical relation between the present work and that of Kolmogoroff. The present investigation was initiated in the early winter of 1940 as an attempt to solve an engineering problem. At that time and until the last

---

[7] The German version is now standard reference in Russell and Norvig's textbook on Artificial Intelligence. See the bibliography session of this textbook.

week of 1941, by which time the paper was substantially complete, the author was not aware of the results of Kolmogoroff's work and scarcely aware of its existence. (1942/1964, 59)

In the 1920s, at least in the USSR, Kolmogorov was already a rising star, having 28 mathematical papers to his credit by the time he finished his four years as a postgraduate. In 1931, he became a professor at Moscow University and was appointed the Director of the Scientific and Research Institute of Mathematics in another two years. He completed his most influential works in the 1930s, including 54 papers and the famous *Foundations of the Theory of Probabilities* (Dudley *et al.* 1990, 1031). Extrapolation occupies an entire chapter in *Selected Works of A.N. Kolmogorov* (Shiryayev 1992). Like Wiener, Kolmogorov had studied statistical theory and "devised a scheme of stochastic distribution of barrage balloons to protect Moscow from Nazi bombers" (Gleick 2011, 335). Also a polymath, he ventured into literature and linguistics while continuing to work in the fast-evolving field of computation. He had worked on similar problems as scientists did in the US (Shiryaev 1989, 888), including "Professor Fuller at Brown University," and was known in some mathematician circles (Shiryaev 1989, 881).

Wiener acknowledges in the 1964 reprint that the overlap was significant. First, both covered ways of filtering, or in Wiener's words "purification" (1942/1964, 9), a function to strip data from the "contamination of other time series" and to maintain a better data collection. Then, Wiener pointed out that the most significant overlap, which was on extrapolation, was a result of different minds working in synchrony - except that his result was more thorough: "Kolmogoroff proceeds only so far as the greatest lower bound of the mean square error of prediction, while we obtain the optimum predicting operators… Thus, it would appear that the work of Kolmogoroff and that of the present writer represent two entirely separate attacks on the problem of time series" (1942/1964, 59). This is the only instance in *Extrapolation* in which Wiener so explicitly tried to control the narrative about the novelty of his work. For Wiener, "the parallelism between them may be attributed to the simple fact that the theory of the stochastic process had advanced to the point where the study of the prediction problem was the next thing on the agenda" (1942/1964, 59). The "next thing on the agenda" that naturally arrived at the great minds in mathematics was this idea of extrapolation, which had become the hidden thread that connected the scientific efforts in WWII and the Cold War.[8]

However, what unfolded in the following years changed Wiener's perception of Kolmogorov. In the many years since the publication of Kolmogorov's work, it had garnered a mere six citations besides Wiener's mention in his report (Dudley *et al.* 1990). It could be attributed to the limited circulation of Soviet sciences, or perhaps the uncanny resemblance between the two mathematicians where citing one was almost akin to invoking the other. Kolmogorov, it seemed, did not command the same attention on this subject as Wiener's reprinted report came out. The proliferation of extrapolation in the field of military strategies also pushed further integration of this concept into other areas of social applications, most notably Jim Simmons' use of it in predicting stock market outcome and the automatic military system developed for the US War on Terror.

**Extrapolation Yes, Gadget No**

In the transformation of human languages into a time series, Wiener thought he was making a leap that was more far-reaching than, say, Claude Shannon's statistical measurement of information content. The competition between Wiener and Shannon is now a well-known tale, but another point related to their different visions is worth discussing. In his famous 1950 speech "Man, Machines, and the World About" addressed to the New York Academy of Medicine and Science, Wiener belittles a certain engineer:

> I know a great engineer who never thinks further than the construction of the gadget and never thinks of the question of the integration between the gadget and human beings in society if we allow things to have a reasonably slow development, then the introduction of the gadget as it naturally comes may hurt us enough to provoke a salutary response. (1954, 26)

---

[8] Seising (2010) also mentioned a similar scene about Wiener's 1948 work, in which Wiener wrote: "Let it be remarked parenthetically that some of my speculations in this direction [of cybernetics] attach themselves to the earlier work of Kolmogorov in Russia, although a considerable part of my work was done before my attention was called to the work of the Russian School" (4461).

Wiener does not name the engineer, but he probably alludes to Shannon. Shannon focuses on formulating efficient logarithmic functions for measuring information and setting coding standards to maximize information transfer within a certain bandwidth. Shannon's primary concern, as Wiener sees it, is to define the amount of information contained in a set of sources that a set of apparatus can handle.[9] In this crucial phase of Wiener's later cybernetics project, Wiener differentiates himself sharply from his younger colleague, whose goal seems mostly to divorce language from its meaning.

In Wiener's characterization, Shannon is a junior engineer obsessed with gadgets. "We cannot worship the gadget and sacrifice the human being to it" (1954, 26). Wiener wants to broaden the field of communication engineering without limiting it to the mere functionality of devices, or the *workings* of gadgets. "The proper field of communication engineering is far wider than that generally assigned to it....the records of current and voltage kept on the instruments of an automatic substation are as truly messages as a telephone conversation…the record of the thickness of a roll of paper kept by a condenser working an automatic stop on a Fourdrinier machine is also a message" (1942/1964, 2). In his view, all problems, if it has to do with service mechanism, all problems can be seen as communication problems. And a communication problem can canvas anything that involves human and organic apparatus.

---

[9] Wiener wrote in a patronizing way when he introduced Shannon's engineering concerns: "Some fifteen years ago, a very bright young student came to the authorities at MIT with an idea for a theory of electric switching dependent on the algebra of logic. The student was Claude E. Shannon." (Gleick 2011, 234).

Kolmogorov, however, is keener in the works of Shannon, saying in contrast that the "mathematical discipline of cybernetics in Wiener's understanding lacks unity" (Gleick 2011, 335). For Kolmogorov, Shannon's attention on communication is focused, or at least less distracted by all kinds of analogies that populate Wiener's work. So, in the 1950s, Kolmogorov became more interested in a similar set of mathematical problems that Shannon was working on, focusing on measuring information in random data and data compression. This line of work eventually led to Kolmogorov's more famous work on data compression.[10]

Wiener's phrase "gadget worshiper" now carries a prophetic significance. In the chatbot goldrush, the "GPT hallucination" emerged as a sign that certain gadgets were working well, and perhaps too well in extrapolating beyond their training data. Were Wiener to witness the present cybernetic embodiment in the chatbot, he would perhaps implore us to look behind an engineer's curtain. "Behold," he might declare, "the chatbot is just another well-functioning contrivance. As I have emphatically said back in 1950, we must be rid of the obsession with the gadget!"

## 3.   Extrapolation, a concept

Since Wiener and Kolmogorov, extrapolation as a concept has taken on significance in many fields adjacent to machine learning (Sidi 2003). Current theories in deep learning ground their theory in Kolmogorov, but these works are mostly focused on compression. So far, I have tried to highlight that extrapolation has been an earlier phase in both Wiener's and Kolgomorov's intellectual enterprise.

---

[10] In 1966, the same year when US mathematicians were focused understanding randomness, computability, and information following the works of Shannon and Weaver, Kolmogorov was again a step ahead and published in a Russian journal and became footnotes to works in his American counterparts. This was mentioned in the passing by Gleick (2011, 333).

In the more technical terms of machine learning, extrapolation refers to mathematical functions that map from an input space to an output space. One can compare the output to the input space, which helps establish the numerical relationship between a testing sample and the training dataset. The training set is a finite data set used to train an AI model. The training set forms a "convex hull," a unique minimal closure containing all the set's points. Figuratively, we can think of such closure as a dome structure. If a sample lies outside the dome of a training set, it means the model must extrapolate to produce an output for the sample. Convex hull algorithms have many use. For example, it can help ensure robots and self-driving cars do not collide with objects. An object outside the convex hull means a self-driving car will not hit it.

In deep learning, convex hull maps the feature space of a model's training set. It marks an outer bound, a minimal enclosure of all the points in the training set. When the decision or output remains inside, the model interpolates. When it reaches beyond the minimal enclosure, it extrapolates. If a model extrapolates while limiting its error within a certain margin, it means that the model has learned something and can begin generating new texts. This is where the LLMs come into the picture. Realignments of details that have now surfaced in the fluid speech of human-like chatbots are bound to scraped web data but freed from its constraining facticity due to the capacity to extrapolate.

However, up until recently, deep learning researchers had assumed that machine learning models do not extrapolate (Belkin 2021). By this assumption, they focus on how models are bound to their training data: predictions happen within the range of the training data, i.e. interpolation. It is believed that the critical function of model is interpolation, whereas extrapolation is nonexistent or insignificant (Belkin *et al.* 2019). A recent provocation busted this long-time misconception in machine learning. A team of high-profile scientists (including Le Cun) suggest that high-dimensional models, the most conspicuous of which is the large language model behind chatbots, "always amounts to extrapolation" (Balestriero *et al.* 2021). "Amounting to extrapolation" is a valuable update to the long-standing misconception in computer science, but the "always" here can introduce another problem. If we look at lower dimension models, not all amounts to extrapolation, but only specific types of predictions require it (Cao and Yousefzadeh 2023). The provocation can leave the impression that since all are extrapolation, it must be somehow banal.

So far in this article, I have shown that extrapolation is neither trivial nor banal. In recent years, there has been increasing attention in the computer science community on the techniques of dataset partitioning and algorithmic encoders to improve extrapolation capacities or evaluate its predictive performances (Zhan *et al.* 2022; Huo *et al.* 2023). Computer science has broadly assumed that the integration of discrete data points is the norm, and methods to achieve this goal - extrapolation included - are the key to enhancing creativity in artificial intelligence. Extrapolations is a procedure to transform a sequence (Brezinski and Redivo Zaglia 1991), not inherently desirable or undesirable in computational mathematics. The ability to generate new ideas is thus conceived as "capacity" that can be improved with input from other fields, including literature. Machine learning scientists have recruited literary theorists to help them understand how real literature extrapolates and how ChatGPT and other LLMs can improve their capacity to produce new text. Through it, LLMs have learned certain correlations of elements, such as Shakespeare's poetic form, but failing in bringing out the same emotional effects (Van Heerden and Bas 2021). But to AI developers who prioritize the poetic form over the poetic effects, such failures were seen as success. They consider their models *working* if they generate the form that makes textual elements hang like poems.

## 4. GPT Hallucinations

Extrapolation, especially those that might result in false references and fake court reports, leaves us with the impression that something must not be working.[11] But what exactly is not working, and for whom? For those who did not believe that machines could generate human-like dialogues, the initial response to the marvelous chatbot GPT was perhaps a kind of unenthusiastic respect. Many acknowledged that with sufficient training in well-structured data, a chatbot can imitate human language at an uncanny level of accuracy. Then, suspicion set in. Researchers have begun recognizing hallucination as a problem and Microsoft teams have begun developing new procedures to ensure that GPT double-check its answers given to prompts regarding factual questions.

Meanwhile, extrapolation has enabled many Digital Humanities projects in which coders and literary people collaborated to write poems and fiction. Towards these other creative AI projects, we may display a genuine admiration. Yet, towards GPT fabricating references, suspicious scholars may feel the urge to downgrade AI to a much earlier stage, likening it to an imperfect Research Assistant. When we hold such inconsistent attitudes towards AI functionality, we are projecting a naive expectation, what Weatherby and Justie call "iconic interpretation." (Weatherby and Justie 2022) This is an expectation that we want an AI model to communicate and work like humans. GPT hallucination was merely an anthropocentric perception of GPT's string of words with no intended meaning other than what was statistically probable in a sequence.

Ask your chatbot if it is hallucinating. It says, "No, I am not hallucinating. I am a computer program that provides information based on the data I have been trained on and your textual prompt." But even in this answer, it merely presents the statistical probabilities through a series of words.

---

[11] Indeed, an emergent topic recently has been helping machine learn models to perform better by improving extrapolating. These works also leave the impression that extrapolation must be something undesirable. Other scholars have argued that extrapolations are not necessarily bad, but sometimes a function necessary for a model to work.

The question is how critical AI community can collectively make sense of the hallucinatory capacity without glimpsing just the surface of this commodity. I use the word commodity deliberately in a Marxian sense to signal what has been concealed by the polished performance of GPT. The commodity refers to the capacity of chatbots to serve as communication partners for human. This communication skills, as Elena Esposito points out, has nothing to do with its core intelligence. Esposito (2022) hints at an important distinction: what we call artificial intelligence is actually artificial communication. For a commercial chatbot, the commodity features include foremost its communication skills. And if the bot fails to serve as a trustworthy communication partner, it is then perceived as not working.[12] Yet, if we peer into how ChatGPT work, we will be surprised to see that the so-called dysfunctionality is evidence that they are working too well, by design. The fact that they can produce codes and dialogues was a design feature; their core intelligence lies in something else, i.e., their capacity to compress data and extrapolate. The machines know how to extrapolate and create, but in their communication with us, such extrapolation might have fallen into the categories of hallucination.

---

[12] There have been new works on arXiv investigating how much we can trust GPT-4 (often compared to GPT 3.5 or 3) in its ability to produce codes, references, responses to politically sensitive content. This kind of research is ongoing. See Chen et al. (2023) and Wang et al. (2023).